\documentclass{article}
\usepackage{authblk}
\usepackage[utf8]{inputenc}
\usepackage[numbers,sort&compress]{natbib}
\usepackage{xpatch}
\usepackage{amsfonts}
\usepackage{amssymb}
\usepackage{latexsym}
\usepackage[final]{hyperref}
\usepackage[]{enumerate}
\usepackage{verbatim}
\usepackage{graphicx}       
\usepackage{amsmath}
\usepackage{tikz}
\usepackage{cmap}
\usepackage{soul}
\usepackage{comment}
\usepackage{fancybox}
\usepackage{relsize}
\usepackage{pifont}

\newcommand{\R}{\mathbb{R}}

\DeclareMathOperator{\sech}{sech}

\begin{document}

\title{Fully dispersive Boussinesq models with uneven bathymetry}

\author[$\dagger$]{John D. Carter}
\author[$\star$]{Evgueni Dinvay}
\author[$\ddag$]{Henrik Kalisch}

\affil[$\dagger$]{Mathematics Department, Seattle University, \textit{carterj1@seattleu.edu}}
\affil[$\star$]{Department of Mathematics, University of Bergen, \textit{Evgueni.Dinvay@uib.no}}
\affil[$\ddag$]{Department of Mathematics, University of Bergen, \textit{Henrik.Kalisch@uib.no}}




\maketitle

\begin{abstract}
	Three weakly nonlinear but fully dispersive Whitham-Boussinesq systems for uneven bathymetry are studied.  The derivation and discretization of one system is presented.  The numerical solutions of all three are compared with wave gauge measurements from a series of laboratory experiments conducted by Dingemans \cite{Dingemans}.  The results show that although the models are mathematically similar, their accuracy varies dramatically.
\end{abstract}


\section{Introduction}
\setcounter{equation}{0}
In coastal engineering, Boussinesq models are used to
approximate full Euler or Navier-Stokes equations which 
are numerically intractable on large scales.
The main assumptions on the waves to be represented by approximate Boussinesq-type models
are that they be of small amplitude and long wavelength when compared to the undisturbed depth of the fluid.
As explained in \cite{MMS}, classical Boussinesq models
are able to accurately describe waves up to a wavelength-to-depth ratio
of $kh \sim 1.3$, where $k=2 \pi / \lambda$ is the wavenumber, 
$\lambda$ is the wavelength, and $h$ is the local depth.
On the other hand, in many practical applications, it is desirable
to be able to treat shorter waves or waves in deeper water,
and the development of coastal models has long been focused 
on obtaining models allowing closer approximation of waves in deeper water. 

One of the first results in this direction was given in \cite{Witting},
where a KdV equation with improved dispersion properties was found.
In \cite{MMS,MS}, two-dimensional Boussinesq equations with
improved dispersion and bathymetry were put forward.
The dispersion relation was further improved by \cite{Nwogu},
and in many subsequent articles.
Current models are able to treat 
smaller wavelength-to-depth ratios than the traditional Boussinesq models,
up to about $kh \sim 30$ \cite{Brocchini2013, MFW, BOSZ}.
However, one drawback with these high-order systems is that they tend to become
very cumbersome to represent in writing and to implement numerically.
In addition, many numerical fixes are used (sometimes tacitly) because 
the modifications done in order to improve linear dispersion
and treatment of bathymetry sometimes introduce instabilities.

In the present work, we consider a class of fully dispersive Boussinesq systems.  These systems are developed using an idea of Whitham~\cite{WhithamBook} who put forward the original nonlinear fully dispersive equation 
\begin{equation}\label{dimWhitham}
\eta_t + \frac{3}{2}\frac{c_0}{h}\eta\eta_x + \int_{-\infty}^\infty \mathcal{K}_{h}(y) \eta_{x} (x-y) \, dy =0,
\end{equation}
where $h$ is the undisturbed depth of the fluid, $c_0 = \sqrt{g h}$ is the corresponding long-wave speed, and  $g$ is the gravitational acceleration.  The integral kernel $\mathcal{K}_{h}$ is given in terms of the Fourier transform and the linear phase speed $c(\xi)$ by
\begin{equation}\label{symbol}
\mathcal{F} \mathcal{K}_{h} (\xi) = c(\xi) = {\textstyle{ \sqrt{ g \frac{\tanh (h\xi)}{\xi}}}}.
\end{equation}

The convolution can be thought of as a Fourier multiplier operator and \eqref{symbol} represents the Fourier symbol of the operator.  Indeed, using the notation $D=-i\partial_x$, the integral operator can be written in the form $\sqrt{gh \mathcal{K}}$, where
\begin{equation}
	\label{Operator_defs}
	\begin{aligned}
		\mathcal K = \frac{\tanh(hD)}{hD}.
	\end{aligned}
\end{equation} 

The linearization of equation (\ref{dimWhitham}) gives an {\it exact} unidirectional representation of the linear dispersion relation of the free-surface water-wave problem. The fidelity of solutions of this equation has been tested against numerical solutions of the the Euler equations \cite{MKD} and measurements from wave tank experiments \cite{Carter}.  In recent work, the Whitham equation, (\ref{dimWhitham}), has been generalized to systems of equations allowing for bi-directional wave propagation.  Essentially, three different forms of the equations have been put forward \cite{Aceves_Sanchez_Minzoni_Panayotaros, Dinvay, Hur_Pandey}, and these have also been tested against laboratory data from experiments with constant depth, see \cite{Carter}.

We consider the influence of bathymetry,
which is an essential feature from the point of view of coastal engineering.
In fact, the model found in \cite{Aceves_Sanchez_Minzoni_Panayotaros}
already featured nontrivial bathymetry,
but the bathymetric terms were somewhat simplified.
Here, we investigate the bidirectional Whitham-type system 
from \cite{Aceves_Sanchez_Minzoni_Panayotaros}
with full bathymetric terms as well as capillarity in the form
\begin{equation}
	\label{ASMP_sys}
	\begin{aligned}
		\partial_t \eta &=
		- h \mathcal{K}\partial_x (1 + \frac{\tau}{g} D^2) u - \partial_x (\eta u)
		- \partial_x L(\beta) D^{-1} u
		, \\
		\partial_t u &=
		- g \partial_x \eta
		- \partial_x u^2 / 2
		,
	\end{aligned}
\end{equation}
where the parameter $\tau$ represents the coefficient of surface tension.  The unknowns $\eta$ and $u$ are real-valued functions of the spatial variable $x\in\mathbb R$ 
and the temporal variable $t \in[0,\infty)$.  
They represent the surface displacement of the fluid and the 
horizontal velocity component defined by
\(
	u = \partial_x \Phi 
\)
where $\Phi(x,t)$ is the surface trace of the fluid's velocity potential $\phi(x, z, t)$.  
The bathymetry operator $L(\beta)$ was introduced in \cite{Craig_Guyenne_Nicholls_Sulem} 
and has the form
\begin{equation}
\label{L_definition}
	L(\beta) = - C(\beta) ^{-1} A(\beta),
\end{equation}
where the operators $A(\beta)$ and $C(\beta)$ are defined by
\begin{equation}
\label{A_definition}
	A(\beta) f = \int e^{ikx} \sinh \big{(}\beta(x) k\big{)}
	\sech(hk) \widehat{f}(k) dk
	= \int e^{ik(x-s)} \frac{ \sinh \big{(}\beta(x) k\big{)} }{ \cosh(hk) }
	f(s) ds dk
	,
\end{equation}
\begin{equation}
\label{C_definition}
	C(\beta) f = \int e^{ikx} \cosh \big{(} (-h + \beta(x)) k \big{)}
	\widehat{f}(k) dk
	= \int e^{ik(x-s)} \cosh \big{(} (-h + \beta(x)) k \big{)}
	f(s) ds dk
	.
\end{equation}
If the bottom is flat, then $\beta(x)=0$ and $L(\beta)=0$, and in this case, 
it was proven in \cite{PeiWang}
that this system is well-posed 
if the surface displacement is strictly positive.  
Additionally, it was suggested that this system is ill-posed in general.  
Numerical results corroborating these well-posedness/ill-posedness statements
were detailed in \cite{ClaassenJohnson, Dinvay_Dutykh_Kalisch}. 
However, as will be shown below, these result do not seem to have a bearing on
the numerical experiments presented here.
Indeed, it was shown in \cite{Carter}
that the flat-bottom version of this system
represents a more accurate model of the evolution of initial waves of depression over 
a flat bottom than the KdV equation and even the higher-order Serre-Green-Naghdi system.

The system has the conserved Hamiltonian function
\begin{equation}
	\label{ASMP_Hamiltonian}
		\mathcal H  = \frac 12 \int_\R
		\Big{(}
			g \eta^2 + hu \mathcal K (1 + \frac{\tau}{g} D^2) u
			+ u L D^{-1} u
			+ \eta u^2
		\Big{)}
		dx
		,
	\end{equation}
and in terms of $\mathcal{H}$ , the system can be written in the form
\begin{equation}
\label{ASMP_Hamiltonian_structure}
\begin{aligned}
	\eta_t = - \partial_x \frac{\delta \mathcal H}{\delta u} 
	, \\
	u_t = - \partial_x \frac{\delta \mathcal H}{\delta \eta}
	.
\end{aligned}
\end{equation}
It was shown in \cite{MKD} that (\ref{ASMP_Hamiltonian}) can 
be viewed as a fully-dispersive approximation of the total energy
of the full water-wave problem.

A different model system was proposed 
in the case of a flat-bottom in \cite{Hur_Pandey}.
In the presence of bathymetry and capillarity,
the system has the form
\begin{equation}
\label{Hur_sys}
\begin{aligned}
	\partial_t \eta &=
	- h \partial_x u - \partial_x (\eta u)
	- \partial_x L(\beta) D^{-1} u
	, \\
	\partial_t u &=
	- g \mathcal K \partial_x (1 + \frac{\tau}{g} D^2) \eta
	- \partial_x u^2 / 2
	,
\end{aligned}
\end{equation} 
where $\eta$ is the free-surface displacement, and $u=\partial_x\Phi(x,t)$
is the horizontal velocity component as before.
The system (\ref{Hur_sys}) also has a Hamiltonian structure.
Indeed, the system can be written in the form
\begin{equation}
\label{Hur_Hamiltonian_structure}
\begin{aligned}
	\eta_t = - \partial_x \frac{\delta \mathcal H}{\delta u} 
	, \\
	u_t = - \partial_x \frac{\delta \mathcal H}{\delta \eta}
	,
\end{aligned}
\end{equation}
with the Hamiltonian
\begin{equation}
\label{Hur_Hamiltonian}
	\mathcal H  = \frac 12 \int_\R
	\Big{(}
		g \eta \mathcal K (1 + \frac{\tau}{g} D^2) \eta + hu^{2}
		+ u L D^{-1} u
		+ \eta u^2
	\Big{)}
	dx
	.
\end{equation}
However this Hamiltonian is not an approximation of the Hamiltonian
of the water-wave problem in the context of the Craig-Sulem-Zakharov formulation
(see for example\cite{Lannes}).  
It was shown in \cite{Hur_Pandey}
that periodic traveling-wave solutions are spectrally unstable with respect to long-wave perturbations
due to the modulational instability.

A third model can be obtained by imposing the operator $\mathcal K$ 
also on the nonlinear parts of \eqref{Hur_sys}.  
This gives the system
\begin{equation}
\label{Regularised_sys}
\begin{aligned}
	\partial_t \eta &=
	- h \partial_x v - \mathcal K \partial_x (\eta v)
	- \partial_x L(\beta) D^{-1} \mathcal K^{-1} v
	, \\
	\partial_t v &=
	- g \mathcal K \partial_x (1 + \frac{\tau}{g} D^2) \eta
	- \mathcal K \partial_x v^2 / 2.
	\end{aligned}
\end{equation}
The unknowns $\eta$ and $v$ are real-valued functions of the spatial variable $x\in\mathbb R$ 
and the temporal variable $t \in[0,\infty)$.  
They represent the surface displacement of the fluid and the ``velocity'' defined by
\(
	v = \mathcal K \partial_x \Phi 
\)
where $\Phi(x,t)$ is the surface trace of the fluid's velocity potential $\phi(x, z, t)$.  
The system given in equation \eqref{Regularised_sys} is a conservative Hamiltonian system.  
In $\eta, v$ variables the Hamiltonian functional, $\mathcal H(\eta, v)$, has the form
\begin{equation}
\label{Reg_Hamiltonian}
	\mathcal H  = \frac 12 \int_\R
	\Big{(}
		g \eta (1 + \frac{\tau}{g} D^2) \eta + hv \mathcal K^{-1} v
		+ v \mathcal K^{-1} L D^{-1} \mathcal K^{-1} v
		+ \eta v^2
	\Big{)}
	dx,
\end{equation}
with the structure map
\[
	J_{\eta,v}
	=
	\begin{pmatrix}
		0 & - \mathcal K \partial_x
		\\
		- \mathcal K  \partial_x & 0
	\end{pmatrix}.
\]
Thus, the Hamiltonian system is given by
\begin{equation}
\label{Hamiltonian_structure}
	\begin{aligned}
	\eta_t = - \mathcal K \partial_x \frac{\delta \mathcal H}{\delta v} 
	, \\
	v_t = - \mathcal K \partial_x \frac{\delta \mathcal H}{\delta \eta}
	.
	\end{aligned}
\end{equation}
The system \eqref{Regularised_sys} 
was first introduced in \cite{Dinvay_Dutykh_Kalisch} in the 
context of an even bed and without capillarity,
and it was shown in \cite{Dinvay, Dinvay_Tesfahun} that in this simpler case,
the system is mathematically well-posed. 
It was also shown
that the simplified system admits solitary-wave solutions \cite{Dinvay_Nilsson}.

The three systems detailed above are similar, yet have very different mathematical
properties. In this article, we aim to study these systems from a modeling point of view,
in order to determine which of these systems holds most promise as a water-wave model.
We start by giving a derivation of \eqref{Regularised_sys} from the full water wave problem by applying the Hamiltonian long-wave approximation presented in \cite{Craig_Groves}.
In fact, the derivation of these three systems is similar, and we chose to show the
derivation of \eqref{Regularised_sys} because it has appeared most recently in the literature.
The derivation is based on an approximation of the Hamiltonian
which approximates the total energy of the water-wave problem based 
on the full Euler equations.

\section{Hamiltonian formulation}
\setcounter{equation}{0}

Consider an inviscid, incompressible, and irrotational fluid with domain $x \in \mathbb R$, $- h + \beta(x) < z < \eta(x, t)$.  Its motion is described by the Laplace equation
\[
	\partial_x^2 \phi + \partial_z^2 \phi = 0,
\]
in the fluid domain, the Neumann boundary condition
\[
	\partial_n \phi = 0,
\]
at the bottom, $z = - h + \beta(x)$,
indicating the fact that the bottom is impenetrable, the kinematic condition
\[
	\partial_t \eta + ( \partial_x \eta ) \partial_x \phi
	- \partial_z \phi = 0,
\]
at the free surface, $z = \eta(x,t)$,
and the Bernoulli equation including surface tension
\[
	\partial_t \phi + \frac 12
	|\nabla \phi|^2 + g \eta
	- \tau
	\partial_x
	\Big(
		\frac
		{ \partial_x \eta }
		{ \sqrt{ 1 + ( \partial_x \eta )^2 } }
	\Big)
	= 0,
\]
also at the free surface.

In order to reduce this system, we introduce the trace of the velocity potential 
at the free surface, $\Phi(x, t) = \phi(x, \eta(x, t), t)$, 
and the Dirichlet-Neumann operator, $G(\eta, \beta)$, via the formula
\begin{equation}
\label{DN_operator_definition}
	G(\eta, \beta) \Phi = \sqrt{ 1 + (\partial_x \eta)^2 }
	\partial_n \phi, 
\end{equation}
where $\partial_n \phi$ is the projection of the surface fluid velocity on the outward normal vector.  
For a more detailed definition of $G(\eta, \beta)$ taking into account 
the appropriate asymptotic conditions on $\phi$, we refer the reader 
to \cite{Alazard_Burq_Zuily,Lannes}.  
Using the Dirichlet-Neumann operator, the full problem reduces to
\begin{equation}
\label{water_wave_sys}
	\begin{aligned}
		\partial_t \eta & = G(\eta, \beta) \Phi
	, \\
	\partial_t \Phi & = - g\eta
	+ \tau
	\partial_x
	\Big(
		\frac
		{ \partial_x \eta }
		{ \sqrt{ 1 + ( \partial_x \eta )^2 } }
	\Big)
	- \frac 12 (\partial_x \Phi)^2
	+ \frac
	{ ( (\partial_x \eta) \partial_x \Phi + G(\eta, \beta) \Phi )^2 }
	{ 2( 1 + (\partial_x \eta)^2 ) },
	\end{aligned}
\end{equation}
posed on the free surface.  A pair $(\eta, \Phi)$ that solves system \eqref{water_wave_sys} 
completely describes the surface waves.  A drawback of this formulation 
is that the Dirichlet-Neumann operator implicitly depends on the surface elevation $\eta$.  
Zakharov~\cite{Zakharov} showed that system \eqref{water_wave_sys} has the Hamiltonian structure
\begin{equation}
\label{Zakharov_Hamiltonian_system}
\begin{aligned}
	\partial_t \eta = \frac{ \delta \mathcal H }{\delta \Phi}
	, \\
	\partial_t \Phi = - \frac{ \delta \mathcal H }{\delta \eta},
	\end{aligned}
\end{equation}
with total energy
\begin{equation}
\label{Zakharov_Hamiltonian}
	\mathcal H(\eta, \Phi) = \frac 12 \int_{\mathbb R}
	\Big(
		g\eta^2 + \Phi G(\eta, \beta) \Phi
		+
		{\textstyle \frac
		{ 2 \tau (\partial_x \eta)^2 }
		{ 1 + \sqrt{ 1 + (\partial_x \eta)^2 } }}
	\Big)
	dx,
\end{equation}
serving as the Hamiltonian.  The first term in the integral, 
which we denote $\mathcal H_p$, represents the potential energy, 
the second term, $\mathcal H_k$, represents the kinetic energy, 
and the last term, $\mathcal H_{\tau}$, represents the capillary energy.  
The surface water wave problem can be further simplified by approximating the Dirichlet-Neumann operator using different explicit expressions.
We also note that there are different Hamiltonian formulations for this problem,
such as detailed for example in \cite{papoutsellis}

\section{Derivation}
\setcounter{equation}{0}

It is well known that the Dirichlet-Neumann operator can be expanded in a power series in $\eta$, 
see for example, \cite{Craig_Guyenne_Nicholls_Sulem}.  
In the weakly nonlinear framework considered here,
we keep the first two terms in this power series, 
and disregard all higher-order terms.  
In other words, we make the approximation $G \approx G^{0} + G^{1}$, 
where
\begin{equation}
	\begin{aligned}
		G^{0} = D \tanh(hD) + DL
		, \\
		G^{1} = D \eta D - G^{0} \eta G^{0}.
	\end{aligned}
\end{equation}
Recall that the operator $L = L(\beta)$ defined above in equation \eqref{L_definition} 
represents bathymetric effects.  
Note also that $G^{0} = G(0, \beta)$ is symmetric, 
i.e.~for any real-valued functions $f_1$ and $f_2$ belonging to the domain 
of $G^{0}$ the following identity
\[
	\int_{\mathbb R} (G^{0} f_1)(x) f_2(x) dx
	=
	\int_{\mathbb R} f_1(x) (G^{0} f_2)(x) dx,
\]
holds.  This follows from the definition of the Dirichlet-Neumann operator and Green's formula.  
In particular, $DL$ is symmetric and therefore $LD^{-1}$ is also symmetric.  
This fact will figure into the analysis below.

In order to 
simplify the system, we introduce four nondimensional parameters: 
$\varepsilon = a_s / h$, $\mu = h^2 / \lambda^2$, $\gamma = a_b / h$, and $\varkappa=\tau/(gh^2)$ that measure nonlinearity, shallowness, bathymetric variation, and capillarity, respectively.  
Here $a_s$ represents a characteristic surface amplitude, $a_b$ represents 
a characteristic bathymetric variation, 
and $\lambda$ represents a characteristic surface wavelength.  
We assume $\mu \ll 1$.  Generally in the Boussinesq regime, $\varepsilon = \mathcal O(\mu)$. 
However below, it is sufficient to assume only $\epsilon=o(1)$.  
Additionally, we assume that the bathymetric variation does not have to be small 
by allowing $\gamma = \mathcal O(1)$.

Linear theory suggests defining $t_0 = \lambda / \sqrt{gh}$ 
and $\Phi_0 = a_s \lambda \sqrt{gh} / h$ to be the units for time and velocity potential.  Therefore, let 
$\tilde x = x / \lambda$,
$\tilde \eta = \eta / a_s$,
$\tilde \beta = \beta / a_b$,
$\tilde t = t / t_0$ and
$\tilde \Phi = \Phi / \Phi_0$ 
be dimensionless variables.  Similarly, it is convenient to take the units of energy to be
\(
	\mathcal H_0 = g a_s^2 \lambda.
\)
The dimensionless Dirichlet-Neumann operator, $G_{\mu}$, is defined by
\[
	G_{\mu} \left( \varepsilon \tilde \eta, \gamma \tilde \beta \right)
	\tilde \Phi
	=
	\frac h{\Phi_0}  G(\eta, \beta) \Phi
	,
\]
and in particular,
\(
	G_{\mu}^0 = 
	G_{\mu} \left( 0, \gamma \tilde \beta \right)
	.
\)
See \cite{Lannes} for a rigorous proof that
\(
	G_{\mu}^0 \tilde \Phi = \mathcal O(\mu).
\)

In dimensionless variables, the operator $\mathcal K$ is written in the form
\[
	\mathcal K = \frac
	{ \tanh(\sqrt{\mu} \tilde D) }{ \sqrt{\mu} \tilde D },
\]
where $\tilde D = -i \partial_{\tilde x}$ is the derivative with respect to the nondimensional horizontal variable $\tilde x$.  Finally, the dimensionless velocity is 
\(
	\tilde v = \mathcal K \partial_{\tilde x} \tilde \Phi
\)
and therefore
\(
	v = \varepsilon \sqrt{gh} \tilde v
	.
\)

The kinetic energy is approximated by
\[
	\mathcal H_k = \frac 12 \int_{\mathbb R}
	\Phi \big{(}G^0 + G^1\big{)} \Phi dx,
\]
where the $G^0$ part is given by
\[
	\int_{\mathbb R} \Phi G^0 \Phi dx
	=
	\int_\R
	\left(
		hv \mathcal K^{-1} v
		+ v \mathcal K^{-1} L D^{-1} \mathcal K^{-1} v
	\right)
	dx
	,
\]
and the $G^1$ part is given by
\[
	\int_{\mathbb R} \Phi G^1 \Phi dx
	=
	\int_{\mathbb R} \Phi D( \eta D \Phi ) dx
	-
	\int_{\mathbb R} \Phi G^0 \left( \eta G^0 \Phi \right) dx
	=
	\int_{\mathbb R} \eta ( \partial_x \Phi )^2 dx
	-
	\int_{\mathbb R} \eta \left( G^0 \Phi \right)^2 dx,
\]
where we have integrated by parts in the first integral and used symmetry property of $G^0$ in the second.  Converting to nondimensional variables gives
\[
	\int_{\mathbb R} \Phi G^1 \Phi dx
	=
	\varepsilon \mathcal H_0
	\left[
		\int_{\mathbb R} \tilde \eta
		\left( \partial_{\tilde x} \tilde \Phi \right)^2 d\tilde x
		-
		\frac 1{\mu}
		\int_{\mathbb R} \tilde \eta
		\left( G_{\mu}^0 \tilde \Phi \right)^2 d\tilde x
	\right]
	=
	\varepsilon \mathcal H_0
	\int_{\mathbb R} \tilde \eta
	\left[
		\left( \mathcal K^{-1} \tilde v \right)^2
		+ \mathcal O(\mu)
	\right]
	d\tilde x
	.
\]
Making use of the small-$\mu$ Taylor expansion, $\mathcal K^{-1} = 1 + \mathcal O(\mu)$, gives
\[
	\int_{\mathbb R} \Phi G^1 \Phi dx
	=
	\varepsilon \mathcal H_0
	\int_{\mathbb R} \tilde \eta \tilde v ^2
	d\tilde x
	\left(
		1 + \mathcal O(\mu)
	\right)
	.
\]
The error term
\(
	\mathcal H_0 \mathcal O(\varepsilon \mu)
\)
is neglected below.  The surface tension energy
\[
	\mathcal H_{\tau}
	=
	\varkappa \mu \mathcal H_0 \int_{\mathbb R}
	\frac
	{ (\partial_{\tilde x} \tilde \eta)^2 }
	{
		1 + \sqrt
		{ 1 + \mu \varepsilon^2 (\partial_{\tilde x} \tilde \eta)^2 } 
	}
	d \tilde x
	=
	\frac{ \varkappa \mu \mathcal H_0 }2 \int_{\mathbb R}
	(\partial_{\tilde x} \tilde \eta)^2
	d \tilde x
	\left(
		1 + \mathcal O \left( \mu \varepsilon^2 \right)
	\right),
\]
where the error term is negligible.  
Note that the linearization of system \eqref{water_wave_sys} has energy equal to 
\(
	\mathcal H_0 \mathcal O(1)
	.
\)
We have also neglected
\(
	\mathcal H_0 \mathcal O( \varepsilon^2 )
\)
in equation \eqref{Zakharov_Hamiltonian} by discarding the high-order terms, $G^n$ with $n \geqslant 2$, in the expansion of the Dirichlet-Neumann operator.  In total, discarding the terms
\(
	\mathcal H_0 \mathcal O( \mu \varepsilon + \varepsilon^2 )
\)
in equation \eqref{Zakharov_Hamiltonian} and converting back to the original dimensional variables leads to the Hamiltonian given in equation \eqref{Reg_Hamiltonian}.  Calculating variational derivatives in \eqref{Hamiltonian_structure} with $\mathcal H$ given by \eqref{Reg_Hamiltonian} gives system \eqref{Regularised_sys}.

\section{Numerical evaluation}
\setcounter{equation}{0}

Define the Fourier transform of a function $f(x)$ by
\[
	\mathcal F\big{(}f(x)\big{)}= \hat{f}(k) = 
	\int e^{-ikx} f(x) dx,
\]
and the inverse Fourier transform of a function $\hat{f}(k)$ by
\[
	\mathcal F^{-1}\big{(}\hat{f}(k)\big{)}= f(x) = 
	\frac{1}{2 \pi} \int e^{ikx} \hat{f}(k) dk.
\]

Any differential operator $\varphi(D)$ can be calculated by
\[
	\varphi(D) =
	\mathcal F^{-1} \varphi \mathcal F,
\]
where $\varphi$ is the operation of multiplication by the function $\varphi$ in Fourier space.
Bathymetric effects are defined by the operator
$ -\partial_x LD^{-1}  = -i DLD^{-1}$.
As with other differential operators,
this one can be calculated as follows
(omitting $-i$ for simplicity)
\[
	DL(\beta)D^{-1}
	=
	\mathcal F^{-1}
	\mathcal Q(\beta)
	\mathcal F,
\]
where
\[
	\mathcal Q(\beta) =
	\mathcal F DL(\beta)D^{-1} \mathcal F ^{-1}
	=
	\left( \mathcal F D \mathcal F ^{-1} \right)
	\mathcal F L(\beta)D^{-1} \mathcal F ^{-1}
	.
\]
Note that
\[
	\mathcal F LD^{-1} \mathcal F ^{-1} =
	- ( C \mathcal F ^{-1} )^{-1} AD^{-1} \mathcal F ^{-1},
\]
where operator $AD^{-1} \mathcal F ^{-1}$
is defined on functions in spectral space
\begin{equation}
\label{mathcal_A}
	\big( \mathcal A \hat{f} \big) (x)
	=
	\left( AD^{-1} \mathcal F ^{-1} \hat{f} \right) (x)
	=
	\int e^{ikx} \frac{ \sinh (\beta(x) k) }{ k \cosh(hk) }
	\hat{f}(k) dk,
\end{equation}
and operator $C \mathcal F ^{-1}$ can be represented as multiplication of operators
\[
	C \mathcal F ^{-1}
	=
	C \sech(hD) \mathcal F ^{-1} \mathcal F \cosh(hD) \mathcal F ^{-1},
\]
with
\begin{equation}
\label{mathcal_C}
	\big( \mathcal C \hat{f} \big) (x)
	=
	\left( C \sech(hD) \mathcal F ^{-1} \hat{f} \right) (x)
	=
	\int e^{ikx} \frac{\cosh ( (-h + \beta(x)) k )}{\cosh(hk)}
	\hat{f}(k) dk
	.
\end{equation}
The last factorization helps to diminish significantly the condition number of the corresponding discretization.  And so,
\[
	( C \mathcal F ^{-1} )^{-1}
	=
	\mathcal F \sech(hD) \mathcal F ^{-1} \mathcal C ^{-1}.
\]
Combining all of this together, gives the following factorization of the bathymetry operator
\[
	-i DL(\beta)D^{-1}
	=
	-i \mathcal F ^{-1}
	\mathcal Q(\beta)
	\mathcal F
	=
	i
	\mathcal F ^{-1}
	\left(
		\mathcal F D \sech(hD) \mathcal F ^{-1}
	\right)
		\mathcal C ^{-1} \mathcal A
	\mathcal F
	,
\]
with $\mathcal A$ and $\mathcal C$ defined by
\eqref{mathcal_A} and \eqref{mathcal_C} respectively.
In system \eqref{Regularised_sys} one applies the operator
$\mathcal K^{-1}$ first, and then the bathymetry operator.

\subsection{Direct discretization of Bathymetry operator.}

Let $\mathbb L$ be the period for the periodic approximation of the problem.  Short waves do not ``feel'' the bottom, see \cite{Andrade_Nachbin2018}.  If additionally one can show that the bathymetry does not cause the creation of short waves, then waves with frequencies $|k| \geqslant \pi M / \mathbb L$ belong to the kernel of the operator $L(\beta) D^{-1}$ for some large enough even integer $M$.  This corresponds to assuming that short waves do not play a significant role in the bathymetry terms in equations (\ref{ASMP_sys}),  (\ref{Hur_sys}), and (\ref{Regularised_sys}).  Define the projection onto low frequency waves by 
\(
	P_0 = \mathcal F^{-1}
	\chi_{[-\pi M / \mathbb L, \pi M / \mathbb L]} \mathcal F
\)
with $\chi$ standing for the indicator function.  The velocity in equation \eqref{Regularised_sys} can be represented as $v = P_0v + (1 - P_0)v$, where the last term belongs to the kernel of the operator $L(\beta) D^{-1}$.  As a result one can write the approximation
\begin{multline*}
	-i DL(\beta)D^{-1} \mathcal K^{-1}
	=
	-i P_0 DL(\beta)D^{-1} \mathcal K^{-1} P_0
	\\
	=
	i \mathcal F ^{-1}
	\chi_{[-\pi M / \mathbb L, \pi M / \mathbb L]}
	\left(
		\mathcal F D \sech(hD) \mathcal F ^{-1}
	\right)
		\mathcal C ^{-1} \mathcal A
	( \mathcal F \mathcal K^{-1} \mathcal F^{-1} )
	\chi_{[-\pi M / \mathbb L, \pi M / \mathbb L]}
	\mathcal F
	.
\end{multline*}
This allows us to replace the operator $\mathcal A$ with
\[
	\big( \mathcal A \hat{f} \big) (x)
	=
	\int _{ -\pi M / \mathbb L }^{ \pi M / \mathbb L }
	e^{ikx} \frac{ \sinh (\beta(x) k) }{ k \cosh(hk) }
	\hat{f}(k) dk,
\]
and the operator $\mathcal C$ the
\[
	\big( \mathcal C \hat{f} \big) (x)
	=
	\int _{ -\pi M / \mathbb L }^{ \pi M / \mathbb L }
	e^{ikx} \frac{\cosh ( (-h + \beta(x)) k )}{\cosh(hk)}
	\hat{f}(k) dk
	.
\]
A discrete approximation of these operators can be obtained
in a manner similar to the discrete Fourier transform on
the grid $ x_l = l \mathbb L / M $
with $l = 0, \ldots, M-1$
and $k_q = 2\pi q /\mathbb L$ with $q = -M/2 + 1, \ldots, M/2$.  
Then the operators $\mathcal A$ and $\mathcal C$ have the form
\[
	\big( \mathcal A \hat{f} \big) (x_l)
	=
	\frac{2\pi}{\mathbb L} \sum _{q = -M/2 + 1}^{M/2} e^{ik_q x_l}
	\frac{ \sinh (\beta(x_l) k_q) }{ k_q \cosh (hk_q) } \hat{f} (k_q),
\]
\[
	\big( \mathcal C \hat{f} \big) (x_l)
	=
	\frac{2\pi}{\mathbb L} \sum _{q = -M/2 + 1}^{M/2} e^{ik_q x_l}
	\frac{\cosh ((-h + \beta(x_l) )k_q)}{ \cosh (hk_q) }	
	\hat{f} (k_q)
	.
\]
Thus the corresponding discrete transforms have the forms
\[
	\mathcal A (l, q)
	=
	\frac{2\pi}{\mathbb L} e^{ik_q x_l}
	\frac{ \sinh (\beta(x_l) k_q) }{ k_q \cosh (hk_q) },
\]
\[
	\mathcal C (l, q)
	=
	\frac{2\pi}{\mathbb L} e^{ik_q x_l}
	\frac{\cosh ((-h + \beta(x_l) )k_q)}{ \cosh (hk_q) }	
	.
\]

For the discretization of the operators like $D$ and $\sech(hD)$, we let $N \geqslant M$ be a power of two and compute $\mathcal F$ and $\mathcal F ^{-1}$ using fast Fourier transforms (FFTs) of dimension $N$.  

\subsection{An alternative evaluation via power series}

The operator $L$ can be represented as a series of the form
\begin{equation}
	\label{LSeries}
	L = \sum _{j=1}^{\infty} L_j,
\end{equation}
	 where
\[
	L_1(\beta) = - \sech(hD) \beta D \sech(hD)
	,
\]
\[
	L_2(\beta) = \sech(hD) \beta D \sinh(hD) L_1,
\]
\[
	L_3(\beta) =
	- \sech(hD) \Big{(}\frac{1}{6}\beta^3D^2-\frac{1}{2}\beta^2D^2\beta+\beta D\tanh hD\beta D\tanh hD\beta\Big{)}D\sech(hD),
\]
and the higher-order terms are given in \cite{Craig_Guyenne_Nicholls_Sulem}.  A similar approach was used in \cite{Aceves_Sanchez_Minzoni_Panayotaros}.

\section{Results and conclusions}
\setcounter{equation}{0}

\begin{figure}
	\begin{center}
		\includegraphics[width=12cm]{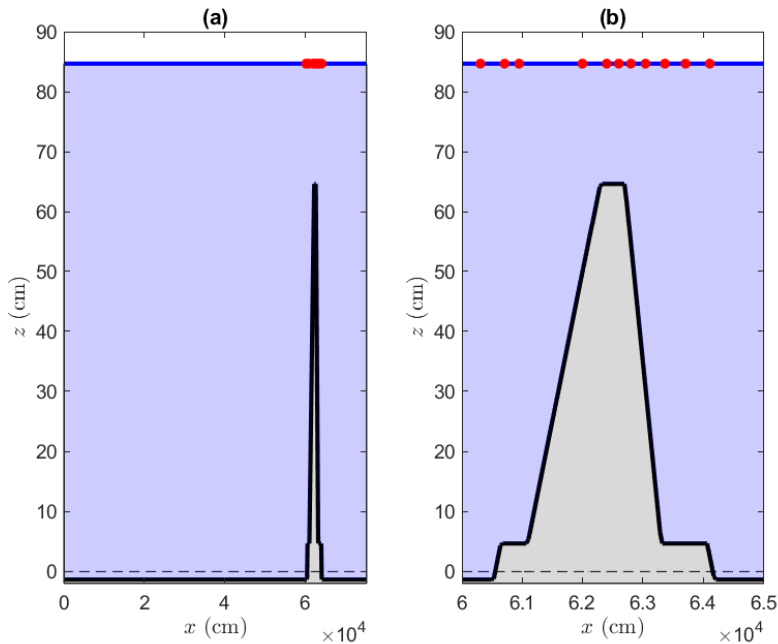}
		\caption{(a) A plot of our entire computational domain.  (b) A zoomed-in plot of the ``seamount.''  
In both plots, the horizontal line at $z\approx 84.68$ is the undisturbed water surface, 
the dashed horizontal line is at $z=0$, and the dots represent the gauge locations.}
		\label{ComputationalDomain}
	\end{center}
\end{figure}

\begin{figure}
	\begin{center}
		\includegraphics[width=16cm]{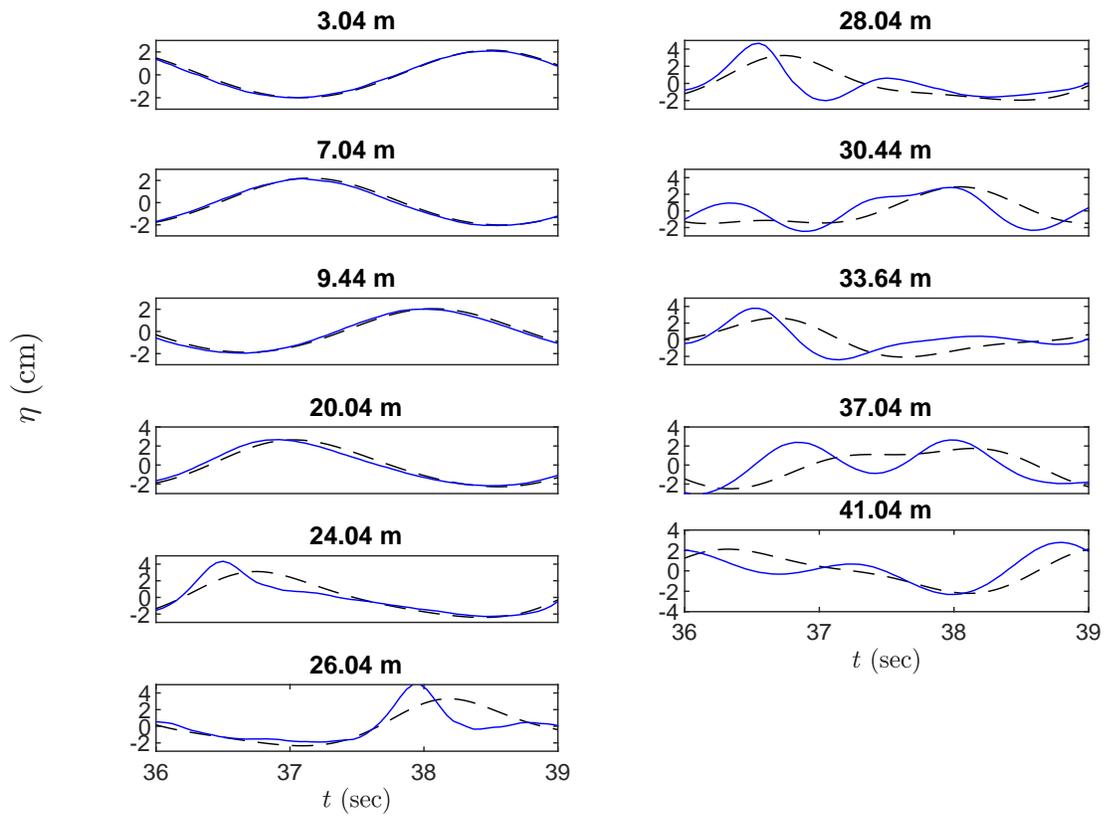}
		\caption{Plots of the experimental time series and the predictions from the system given in equation (\ref{Regularised_sys}) at the eleven gauges ordered by distance from the wave maker.  The solid curves are the experimental time series and the dashed curves are the model's predictions.}
		\label{RegularisedPlot}
	\end{center}
\end{figure}

\begin{figure}
	\begin{center}
		\includegraphics[width=16cm]{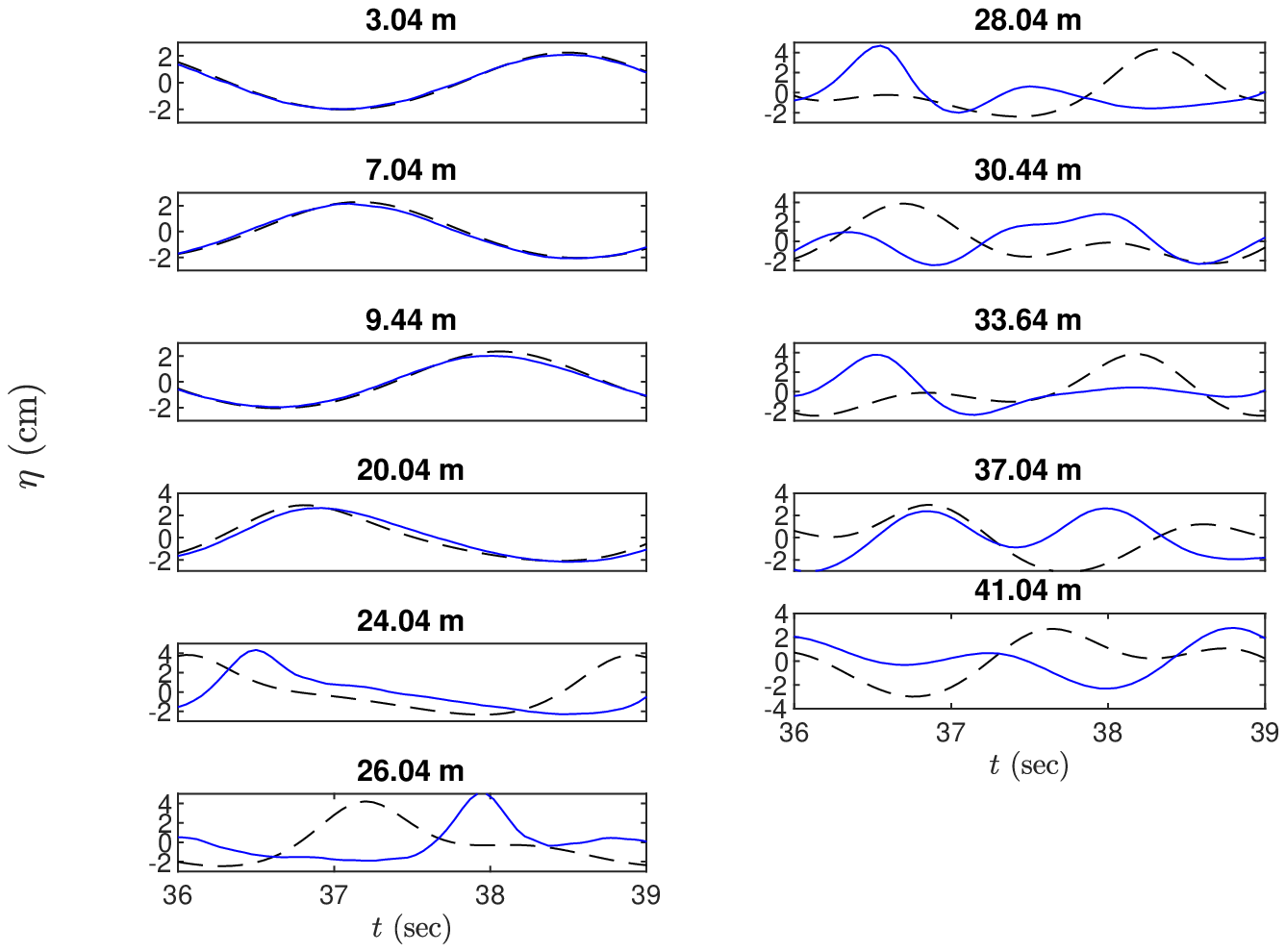}
		\caption{Plots of the experimental time series and the predictions from the system given in equation (\ref{Hur_sys}) at the eleven gauges ordered by distance from the wave maker.  The solid curves are the experimental time series and the dashed curves are the model's predictions.}
		\label{HPPlot}
	\end{center}
\end{figure}

\begin{figure}
	\begin{center}
		\includegraphics[width=16cm]{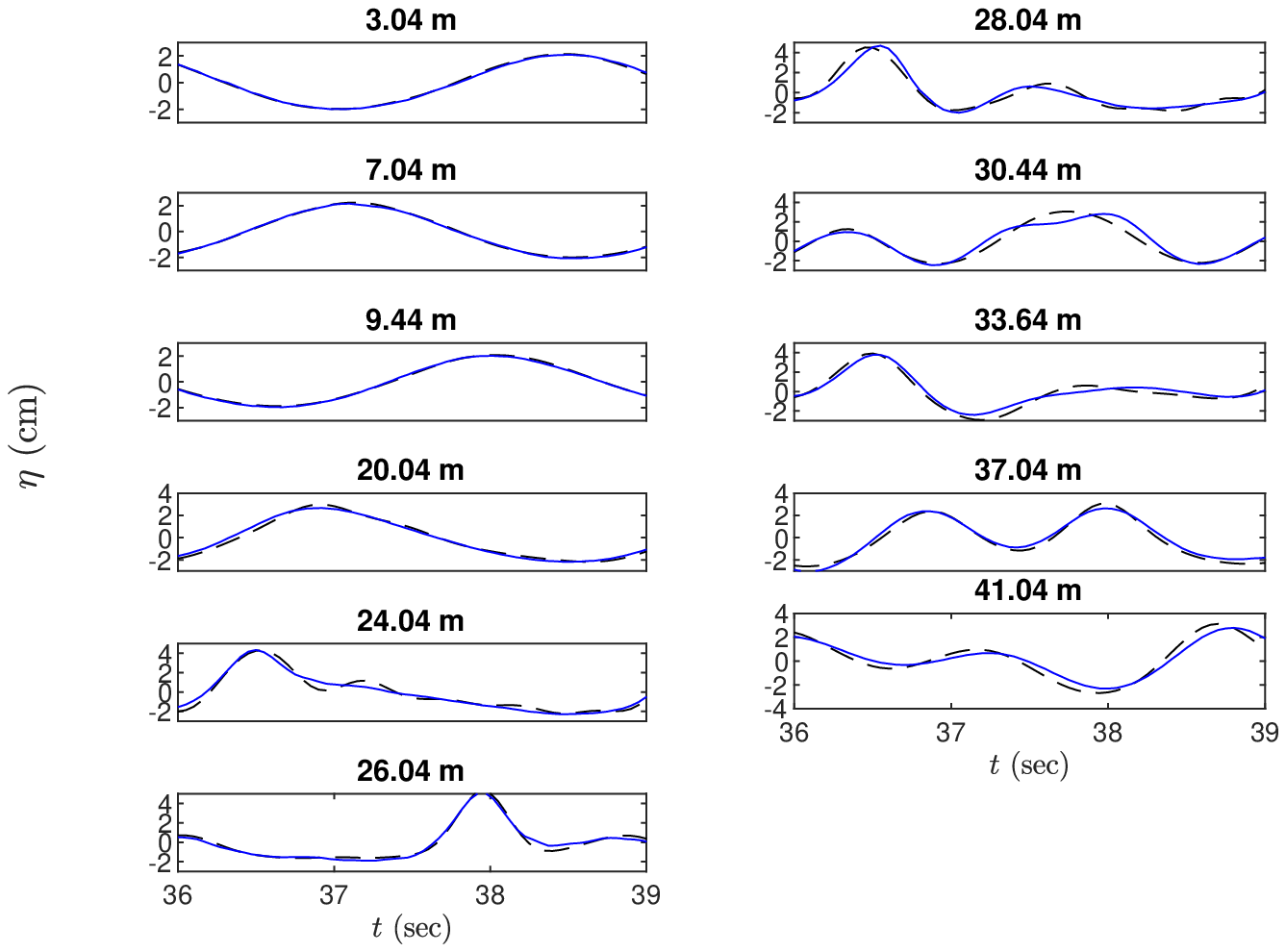}
		\caption{Plots of the experimental time series and the predictions from the system given in equation (\ref{ASMP_sys}) at the eleven gauges ordered by distance from the wave maker.  The solid curves are the experimental time series and the dashed curves are the model's predictions.}	
		\label{ASMPPlot}
	\end{center}
\end{figure}

\begin{figure}
	\begin{center}
		\includegraphics[width=16cm]{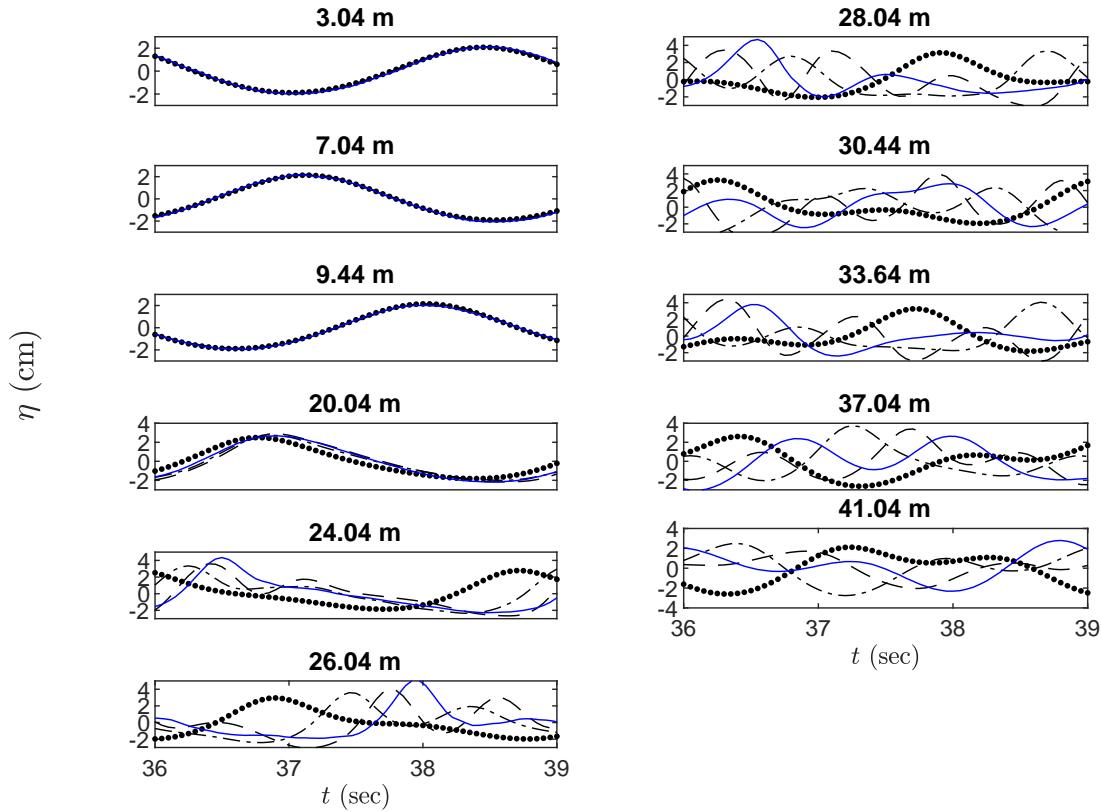}
		\caption{Plots of the experimental time series (solid curve) and the predictions from the system given in equation (\ref{ASMP_sys}) using the $L_1$ approximation (dotted curve), $L_1+L_2$ approximation (dash dotted curve), and the $L_1+L_2+L_3$ approximation (dashed curve).}
		\label{ASMPLsPlot}
	\end{center}
\end{figure}

In order to test the validity and accuracy of these three models, we compare their predictions with the experimental data collected in \cite{Dingemans}.  These experimental measurements have been used as a benchmark for a number of Boussinesq models (see  \cite{Dingemans, GobbiKirby,ChazelLannesMarche}).   In the laboratory experiments, surface water waves were created at one end of a tank by a vertically moving paddle.  These waves traveled down the tank and over a ``seamount'' to the other end of the tank where they were dissipated.  Plots of the bathymetry are shown in Figure \ref{ComputationalDomain}.   Note that our domain is vertically shifted with respect to that used in \cite{Dingemans} because we require the bathymetry to have zero mean. This is simply a choice of coordinates and does not affect the dynamics. The undisturbed water depth over the seamount is $20$ cm.  Eleven wave gauges located near the seamount recorded time series of the free surface deflection as the waves propagated.  Data was recorded every 0.05 seconds.  The time series data for the gauges ordered by distance from the wave maker are included in Figures \ref{RegularisedPlot}-\ref{ASMPLsPlot}.  These plots are discussed in more detail below.

We numerically solve the systems given in equations (\ref{ASMP_sys}),  (\ref{Hur_sys}), and (\ref{Regularised_sys}) using sixth-order operator splitting in time and a Fourier basis in space.  The Fourier basis in space allows the linear, non-bathymetric parts of the models to be solved exactly (to within spectral resolution).  This ensures that the linear phase speed of the waves is accurately reproduced.  No dissipation of any sort, physical or numerical, was included in our codes.  More specifically, we used no filtering to prevent numerical instabilities.  We used $\tau=72.86~\text{cm}^3/\text{s}^2$ and $g=981~\text{cm}/\text{s}^2$ as the coefficient of surface tension and the acceleration due to gravity respectively.  It is important to note that applying $\mathcal{Q}$ is a multiplication of a symmetric matrix in Fourier space.  This means applying the operator $\mathcal{Q}$ is an $\mathcal{O}(M^2/4)$ operation.  Finally, note that the inverse of $\mathcal C$ only needs to be computed once per simulation (not per time step) since the bathymetry does not change in time.

\begin{figure}
	\begin{center}
		\includegraphics[width=12cm]{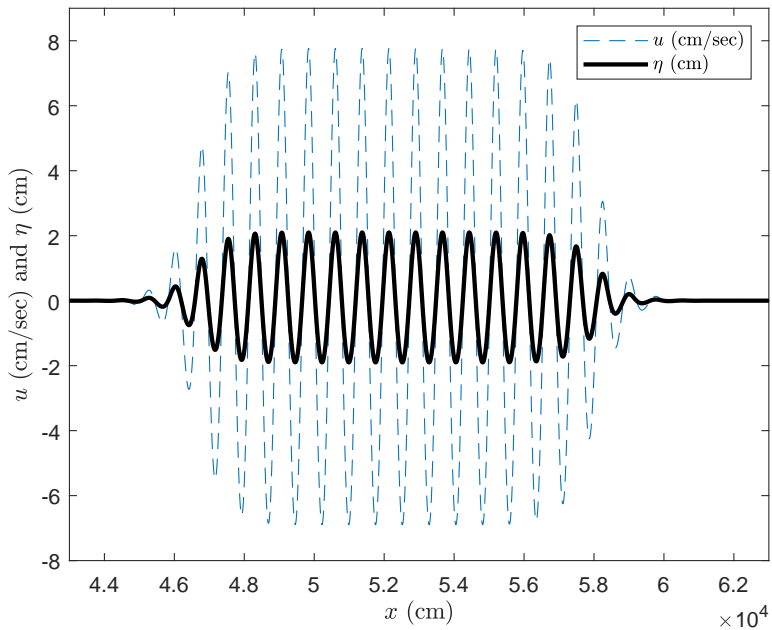}
		\caption{A zoomed-in plot of the initial conditions for the simulations.  The thick solid curve represents the initial surface displacement, $\eta(x,t=0)$, and the thin dashed curve represents the initial horizontal velocity, $u(x,t=0)$.  Both $\eta(x,t=0)$ and $u(x,t=0)$ are zero everywhere outside of this interval.}
		\label{ICs}
	\end{center}
\end{figure}

A drawback of using a Fourier basis is that the motion of the wave maker at the boundary  cannot easily be reproduced. Therefore, we chose the initial conditions to consist of second-order Stokes waves multiplied by an envelope with compact support placed just before the seamount.  The amplitude ($2$ cm), temporal period ($2.86$ sec), and wavelength ($766$ cm) of the waves were chosen to match the experimental wave parameters.  Figure \ref{ICs} includes a plot of the initial conditions.  Since we used periodic boundary conditions and did not include dissipation, we needed to use a computational domain that was large enough that waves did not ``wrap around'' at the right boundary of the domain.  This forced our computational domain to be much larger than the experimental tank.  We used a numerical tank that with length $24000\pi$ cm.  Requiring the bathymetry to have zero mean gives $h\approx84.68$.  A plot of the entire computational domain is shown in Figure \ref{ComputationalDomain}(a).  

All three systems were solved using $N=2048$ (resolution of $\eta$ and $u$ or $v$), $M=2048$ (resolution of the bathymetry), and a time step of $0.05$ seconds.  For the systems given in equations (\ref{Hur_sys}) and (\ref{Regularised_sys}), the conserved quantities (the integral of $\eta$, the integral of $u$ or $v$, and the Hamiltonians) were preserved to within eight or more places.  Increasing the spatial or temporal resolution does not lead to a significant improvement in the preservation of the conserved quantities or a notable difference in the surface displacement predictions.  This is likely due to the fact that the condition number of the matrix $\mathcal{C}$ increases from roughly 200 when $M=N=2048$ to roughly $60,000$ when $M=N=4096$ which leads to more error in computing $\mathcal{C}^{-1}$.

The Hamiltonian for the system given in equation (\ref{ASMP_sys}) was preserved only to three places, while the other two conserved quantities were preserved to within 10 places.  Increasing the spatial resolution to $N=M=4096$ led to numerical instabilities that destroyed the solution before the waves traveled over the entire seamount.  (Recall that our code included no filtering nor dissipation.)  The difficulty preserving the Hamiltonian and the numerical instabilities may be related to the possible ill-posedness of the system.

Figures \ref{RegularisedPlot}, \ref{HPPlot}, and \ref{ASMPPlot} include the results from 
the systems given in equations (\ref{Regularised_sys}), (\ref{Hur_sys}), and (\ref{ASMP_sys}) respectively.  These plots show that system (\ref{ASMP_sys}) provides the best approximation of the experimental data by far.  This may be surprising because this system is thought to be ill-posed when the surface displacement is sometimes negative (as it is here).  The three models provide similar predictions for the surface displacement at the first three gauges (i.e.~before the bathymetry).  They accurately reproduce the experimental measurements at the first three gauge locations.  This suggests that all three models are accurate in the flat-bottom regime.  The model predictions start to deviate at the fourth gauge and this deviation increases as the waves travel over the seamount.  Setting the coefficient of surface tension to zero leads to plots that are indistinguishable to the naked eye from the plots shown.  This suggests that capillarity did not play a significant role in the experiments considered here. However, it is well known that capillarity may be important in different settings (see for example \cite{Aston,RemonatoKalisch}).  Regarding the nonlinear terms, we can report that predictions obtained from the linearized versions of all three models, including system (1.4), provide poor reproductions of the experimental data (plots omitted for brevity).  Indeed, these considerations motivated us to aim for a versatile model which incorporates not only bathymetric effects, but also nonlinearity and capillarity.

Over the time intervals we considered, none of these systems exhibited the oscillatory instabilities 
found in \cite{ClaassenJohnson,Dinvay_Dutykh_Kalisch,MadsenFuhrman}.  Additional simulations (not shown) suggest that there may be a ratio between the amount of negative surface displacement and the length of the computational domain that determines the instability's onset time and/or existence.  It remains unclear why the accuracy of these models varies so much, though it must be said that the model which performs best by far is (\ref{ASMP_sys}) which is given directly in terms of the Hamiltonian structure of the fully nonlinear water-wave problem.  

For comparative purposes, we developed an alternative code
using up to three of the terms in the series approximation to $L$ 
given in equation (\ref{LSeries}).  
These computations are faster than the computations 
of the full systems because the operators $L_j$ can be computed entirely with 
FFTs while applying the complete $L$ operator requires a convolution which takes
$\mathcal{O}(M^2/4)$ operations.
On the other hand, increasing the number of $L_j$ included in the approximation (\ref{LSeries})
also requires a larger number of FFTs in the computation,
so three terms appear
to be a reasonable compromise. Figure \ref{ASMPLsPlot} shows that as the number of terms in the approximation 
of the system given in equation (\ref{ASMP_sys}) increases, the results approach those shown 
in Figure \ref{ASMPPlot}.  The other two models produce similar results.  

To summarize, the system (\ref{ASMP_sys}) which is obtained from a direct
approximation of the Hamiltonian of the water-wave systems performs best
of the three models tested here. The other models are also Hamiltonian,
but use either different canonical variables or a different Hamiltonian
structure, and this is a potential reason why the system (\ref{ASMP_sys})performs best.
One might note that good agreement with the data of Dingemans has been
found by other authors based on Boussinesq codes (see \cite{ChazelLannesMarche} for example).
However, these higher-order models generally have many parameters which need
to be tuned. On the other hand, the model proposed here needs no tuning at all.

For future work, it would be interesting to extend the systems studied here
to more highly nonlinear situations.
Indeed in some cases large wave heights occur due to storms, and especially near the surf zone,
and in this case, Boussinesq models may cease to be applicable. For larger wave heights,
it is possible to use higher-order Boussinesq or Serre-Green-Naghdi models, such
as detailed in \cite{LannesBonneton,WeiKirby}. In order to capture
breaking waves in the surf zone, some models transition to a non-dispersive
shallow-water system based on a breaking parameter, such as explained in
\cite{BacigaluppiETAL,BjKa2011,KazoleaETAL,TonelliPetti,TissierETAL}.
However, so far we do not know of a system combining fully dispersive
properties with a fully or even moderately nonlinear approximation.

\vskip 0.05in
\noindent
\section*{Acknowledgments}

This research was supported by the U.S.~National Science Foundation under grant number DMS-1716120 (JDC), the Research Council of Norway under grant no.~239033/F20 (ED \& HK), and the European Union's {\em Horizon 2020} research and innovation programme under grant agreement No.~763959 (HK).  Additionally, a Fulbright Core Scholar Award allowed JDC to spend a semester visiting HK and ED at the University of Bergen.


\begin{thebibliography}{10}

	\bibitem{Aceves_Sanchez_Minzoni_Panayotaros}
	P.~Aceves-{S{\'a}nchez}, A.A. Minzoni, and P.~Panayotaros.
	\newblock {Numerical study of a nonlocal model for water-waves with variable
	  depth}.
	\newblock {\em Wave Motion}, 50:80--93, 2013.
	
	\bibitem{Alazard_Burq_Zuily}
	T.~Alazard, N.~Burq, and C.~Zuily.
	\newblock {On the water-wave equations with surface tension}.
	\newblock {\em Duke Mathematical Journal}, 158:413--499, 2011.
	
	\bibitem{Andrade_Nachbin2018}
	D.~Andrade and A.~Nachbin.
	\newblock {A three-dimensional {D}irichlet-to-{N}eumann operator for water
	  waves over topography}.
	\newblock {\em Journal of Fluid Mechanics}, 845:321--345, 2018.
	
	\bibitem{Aston}
	P.J. Aston.
	\newblock Local and global aspects of the (1, n) mode interaction for
	  capillary-gravity waves.
	\newblock {\em Physica D}, 52:415--428, 1991.
	
	\bibitem{BacigaluppiETAL}
	P.~Bacigaluppi, M.~Ricchiuto, and P.~Bonneton.
	\newblock Implementation and evaluation of breaking detection criteria for a
	  hybrid {B}oussinesq model.
	\newblock {\em Water Waves}, 2:207--241, 2020.
	
	\bibitem{BjKa2011}
	M.~{Bj\o rkav\aa g} and H.~Kalisch.
	\newblock Wave breaking in {B}oussinesq models for undular bores.
	\newblock {\em Physics Letters A}, 375:1570--1578, 2011.
	
	\bibitem{Brocchini2013}
	M.~Brocchini.
	\newblock {A reasoned overview on {B}oussinesq-type models: the interplay
	  between physics}.
	\newblock {\em Proceedings of the Royal Society of London Series A}, 469:1--27,
	  2013.
	
	\bibitem{Carter}
	J.D. Carter.
	\newblock {Bidirectional {W}hitham equations as models of waves on shallow
	  water}.
	\newblock {\em Wave Motion}, 82:51--61, 2018.
	
	\bibitem{ChazelLannesMarche}
	F.~Chazel, D.~Lannes, and F.~Marche.
	\newblock Numerical simulations of strongly nonlinear and dispersive waves
	  using a {G}reen-{N}aghdi model.
	\newblock {\em Journal of Scientific Computing}, 48:105--116, 2011.
	
	\bibitem{ClaassenJohnson}
	K.M. Claassen and M.A. Johnson.
	\newblock Numerical bifurcation and spectral stability of wavetrains in
	  bidirectional {W}hitham models.
	\newblock {\em Studies in Applied Mathematics}, 141:205--246, 2018.
	
	\bibitem{Craig_Groves}
	W.~Craig and M.D. Groves.
	\newblock {Hamiltonian long-wave approximations to the water-wave problem}.
	\newblock {\em Wave Motion}, 19:367--389, 1994.
	
	\bibitem{Craig_Guyenne_Nicholls_Sulem}
	W.~Craig, P.~Guyenne, D.P. Nicholls, and C.~Sulem.
	\newblock {Hamiltonian long-wave expansions for water waves over a rough
	  bottom}.
	\newblock {\em Proceedings of the Royal Society of London Series A},
	  461:839--873, 2005.
	
	\bibitem{Dingemans}
	M.W. Dingemans.
	\newblock Comparison of computations with {B}oussinesq-like models and
	  laboratory measurements.
	\newblock {\em Delft Hydraulics memo H1684.12}, 1994.
	
	\bibitem{Dinvay}
	E.~Dinvay.
	\newblock {On well-posedness of a dispersive system of the
	  {W}hitham-{B}oussinesq type}.
	\newblock {\em Applied Mathematics Letters}, 88:13--20, 2019.
	
	\bibitem{Dinvay_Dutykh_Kalisch}
	E.~Dinvay, D.~Dutykh, and H.~Kalisch.
	\newblock {A comparative study of bi-directional {W}hitham systems}.
	\newblock {\em Applied Numerical Mathematics}, 141:248--262, 2019.
	
	\bibitem{Dinvay_Nilsson}
	E.~Dinvay and D.~Nilsson.
	\newblock {Solitary wave solutions of a {W}hitham-{B}oussinesq system}.
	\newblock {\em arXiv:1902.09438}, 2019.
	
	\bibitem{Dinvay_Tesfahun}
	E.~Dinvay, S.~Selberg, and A.~Tesfahun.
	\newblock Well-posedness for a dispersive system of the {W}hitham-{B}oussinesq
	  type.
	\newblock {\em SIAM Journal on Mathematical Analysis}, 52:2353--2382, 2020.
	
	\bibitem{GobbiKirby}
	M.F. Gobbi and J.T. Kirby.
	\newblock Wave evolution over submerged sills: tests of a high-order
	  {B}oussinesq model.
	\newblock {\em Coastal Engineering}, 37:57--96, 1999.
	
	\bibitem{Hur_Pandey}
	V.M. Hur and A.K. Pandey.
	\newblock {Modulational instability in a full-dispersion shallow water model}.
	\newblock {\em Studies in Applied Mathematics}, 142:3--47, 2019.
	
	\bibitem{KazoleaETAL}
	M.~Kazolea, A.~Delis, and C.~Synolakis.
	\newblock {Numerical treatment of wave breaking on unstructured finite volume
	  approximations for extended {B}oussinesq-type equations}.
	\newblock {\em Journal of Computational Physics}, 271:281--305, 2014.
	
	\bibitem{Lannes}
	D.~Lannes.
	\newblock {\em {The Water Waves Problem}}.
	\newblock American Mathematical Society, Providence, RI, 2013.
	
	\bibitem{LannesBonneton}
	D.~Lannes and P.~Bonneton.
	\newblock {Derivation of asymptotic two-dimensional time-dependent equations
	  for surface water wave propagation}.
	\newblock {\em Physics of Fluids}, 21:016601, 2009.
	
	\bibitem{MadsenFuhrman}
	P.A. Madsen and D.R. Fuhrman.
	\newblock Trough instabilities in {B}oussinesq formulations for water waves.
	\newblock {\em Journal of Fluid Mechanics}, 889:A38, 2020.
	
	\bibitem{MMS}
	P.A. Madsen, D.R. Fuhrman, and O.R. {S\o rensen}.
	\newblock A new form of the {B}oussinesq equations with improved linear
	  dispersion characteristics.
	\newblock {\em Coastal Engineering}, 15:371--388, 1991.
	
	\bibitem{MFW}
	P.A. Madsen, D.R. Fuhrman, and B.~Wang.
	\newblock A {B}oussinesq-type method for fully nonlinear waves interacting with
	  rapidly varying bathymetry.
	\newblock {\em Coastal Engineering}, 53:487--504, 2006.
	
	\bibitem{MS}
	P.A. Madsen and O.R. {S\o rensen}.
	\newblock A new form of the {B}oussinesq equations with improved linear
	  dispersion characteristics. {P}art 2. {A} slowly-varying bathymetry.
	\newblock {\em Coastal Engineering}, 18:183--204, 1992.
	
	\bibitem{MKD}
	D.~Moldabayev, H.~Kalisch, and D.~Dutykh.
	\newblock The {W}hitham equation as a model for surface water waves.
	\newblock {\em Physica D}, 309:99--107, 2015.
	
	\bibitem{Nwogu}
	O.~Nwogu.
	\newblock Alternative form of {B}oussinesq equations for nearshore wave
	  propagation.
	\newblock {\em Journal of Waterway, Port, Coastal and Ocean Engineering},
	  119:618--638, 1993.
	
	\bibitem{papoutsellis}
	C.E. Papoutsellis.
	\newblock Numerical simulation of non-linear water waves over variable
	  bathymetry.
	\newblock {\em Procedia Computer Science}, 66:174--183, 2015.
	
	\bibitem{PeiWang}
	L.~Pei and Y.~Wang.
	\newblock A note on well-posedness of bidirectional {W}hitham equation.
	\newblock {\em Applied Mathematics Letters}, 98:215--223, 2019.
	
	\bibitem{RemonatoKalisch}
	F.~Remonato and H.~Kalisch.
	\newblock Numerical bifurcation for the capillary {W}hitham equation.
	\newblock {\em Physica D}, 343:51--62, 2017.
	
	\bibitem{BOSZ}
	V.~Roeber, K.F. Cheung, and M.H. Kobayashi.
	\newblock {Shock-capturing {B}oussinesq-type model for nearshore wave
	  processes}.
	\newblock {\em Coastal Engineering}, 57:407--433, 2010.
	
	\bibitem{TissierETAL}
	M.~Tissier, P.~Bonneton, F.~Marche, F.~Chazel, and D.~Lannes.
	\newblock {A new approach to handle wave breaking in fully non-linear
	  {B}oussinesq models}.
	\newblock {\em Coastal Engineering}, 67:54--66, 2012.
	
	\bibitem{TonelliPetti}
	M.~Tonelli and M.~Petti.
	\newblock {Simulation of wave breaking over complex bathymetries by a
	  {B}oussinesq model}.
	\newblock {\em Journal of Hydraulic Research}, 49:473--486, 2011.
	
	\bibitem{WeiKirby}
	G.~Wei, J.T. Kirby, S.T. Grilli, and R.~Subramanya.
	\newblock A fully nonlinear {B}oussinesq model for surface waves. {P}art 1.
	  {H}ighly nonlinear unsteady waves.
	\newblock {\em Journal of Fluid Mechanics}, 294:71--92, 1995.
	
	\bibitem{WhithamBook}
	G.B. Whitham.
	\newblock {\em {Linear and nonlinear waves}}.
	\newblock John Wiley \& Sons, New York, NY, 1974.
	
	\bibitem{Witting}
	J.M. Witting.
	\newblock A unified model for the evolution of nonlinear water waves.
	\newblock {\em Journal of Computational Physics}, 56:203--236, 1984.
	
	\bibitem{Zakharov}
	V.~Zakharov.
	\newblock {Weakly nonlinear waves on the surface of an ideal finite depth
	  fluid}.
	\newblock {\em American Mathematical Society Translations}, 182:167--197, 1998.
	
	\end{thebibliography}

\end{document}